\documentclass[11pt, a4paper]{article}

\usepackage{jheppub}
\usepackage{amsmath, bm, braket, orcidlink}

\begin{document}

\title{
Big Bang Nucleosynthesis as a Probe of First-Order Phase Transitions
}

\author{Masanori Tanaka\,\orcidlink{0000-0002-1303-7043}}

\emailAdd{tanaka@pku.edu.cn}
\affiliation{Center for High Energy Physics, Peking University, Beijing 100871, China}

\abstract{
Supercooled cosmological first-order phase transitions (FOPTs) generate a large temperature inhomogeneity due to their stochastic reheating process. If a baryon asymmetry production predates the FOPT and is conserved during it, the temperature contrast is transferred onto the baryon-to-photon ratio $\eta$ or the baryon-to-entropy ratio $Y_{B}$, producing an inhomogeneity probed by big bang nucleosynthesis (BBN). We extend a recent formalism for this effect by including the reheating dependence of the induced $Y_{B}$ fluctuations, combining the transition-time power spectrum from finite-bubble statistics with proton and neutron diffusion at the BBN epoch. When the nucleation temperature is in the range $10^{-2}\,{\rm GeV} \lesssim T_{n} \lesssim 10^{2}\,{\rm GeV}$, the precise measurement of deuterium abundances disfavors parameter regions with a sizable latent heat and a long transition duration at the $1\sigma$ level. The bound weakens towards higher $T_{n}$ as diffusion becomes more efficient. We further show that this bound disfavors FOPT parameter regions compatible with the NANOGrav gravitational wave signal. This conclusion assumes that the baryon asymmetry predates the transition and that the fluctuation spectrum remains valid down to sub-bubble scales. Therefore, in such FOPT explanations of the NANOGrav signal, the baryon asymmetry cannot simply predate the transition. It can instead be generated after reheating or during the transition if the local baryon production tracks the local entropy injection. As a result, the timing of baryogenesis may be constrained by combining gravitational wave observations with light element abundance measurements.
}

\maketitle

\section{Introduction}

One of the most successful achievements of the standard cosmology is the explanation of the primordial light element abundances through big bang nucleosynthesis (BBN).
In addition, BBN currently provides the earliest credible probe of the thermal history of the early universe.
For instance, the agreement between BBN predictions and the observed light-element abundances constrains the number of additional effective neutrino species $\Delta N_{\rm eff}$~\cite{Planck:2018vyg}, which plays an important role in constraining new physics models with additional relativistic particles in the relatively late universe~\cite{Bai:2021ibt,Yeh:2022heq,Schoneberg:2024ifp}.

It has recently been emphasized that precision measurements of light-element abundances can strongly constrain baryogenesis scenarios that produce a large inhomogeneity in the baryon-to-photon ratio $\eta \equiv n_{B}/n_{\gamma}$~\cite{Bagherian:2025puf, Azatov:2026sdm}.
The effects of spatial inhomogeneities in the baryon-to-photon ratio on BBN have also been studied in Refs.~\cite{Applegate:1987hm, Barrow:2018yyg}.
Since the light-element abundances depend sensitively on the baryon asymmetry of the universe (BAU), precision BBN measurements can, in principle, constrain baryogenesis scenarios directly.
In particular, in several baryogenesis scenarios, the baryon asymmetry is produced with a large initial inhomogeneity.
Conventionally, this inhomogeneity is expected to be smeared by diffusion dynamics.
However, Ref.~\cite{Bagherian:2025puf} showed that the diffusion of such baryon-asymmetry inhomogeneities may not be complete by the time of BBN.
Interestingly, if baryon-asymmetry inhomogeneities produced at temperatures below a few TeV span the horizon scale, they may evade diffusion and survive until the BBN epoch~\cite{Bagherian:2025puf}.
Thus, BBN can potentially carry information about new physics even when the associated energy scale is as high as a few TeV.
Ref.~\cite{Bagherian:2025puf} also pointed out that FOPT interpretations of the pulsar timing array (PTA) signal are natural targets of this constraint.

This constraint applies not only to baryogenesis scenarios that directly produce an inhomogeneous $\eta$, but also to supercooled cosmological first-order phase transitions (FOPTs) because the stochastic timing of bubble nucleation generates a large temperature inhomogeneity through the associated reheating process~\cite{Bagherian:2025puf}.
Therefore, we may be able to explore FOPTs at high energy scales by measuring the abundance of light elements such as deuterium, whose primordial abundance is precisely determined from quasar absorption-line measurements~\cite{Cooke:2017cwo}.

Inspired by this finding, Ref.~\cite{Azatov:2026sdm} discussed BBN-based constraints on FOPT properties.
In particular, precision BBN measurements were shown to exclude $\beta/H \lesssim 2 \text{--} 8$, where $\beta$ and $H$ denote the inverse duration of the phase transition and the Hubble parameter, respectively.
While that analysis accounted for temperature fluctuations arising from differences in the local phase-transition time, it did not include the effect of reheating on those fluctuations.
In general, supercooled FOPTs with a small value of $\beta/H$ release a large latent heat, so the reheating effect also plays an important role in constraining FOPTs from BBN measurements.

Other constraints on supercooled FOPTs with small $\beta/H$ have been derived from curvature perturbations relevant to BBN~\cite{Liu:2022lvz}, which exclude the region of large latent heat parameterized by $\alpha_n$.
This further underscores the importance of the reheating effect for constraints on FOPTs from baryon-asymmetry inhomogeneities.
Constraints from primordial black hole (PBH) observations have also been discussed~\cite{Liu:2021svg,Hashino:2021qoq,Kawana:2022olo,Hashino:2022tcs,Lewicki:2023ioy,Gouttenoire:2023naa,Gouttenoire:2023bqy,Lewicki:2024ghw,Kanemura:2024pae,Cai:2024nln,Florentino:2024kkf,Hashino:2025fse,Franciolini:2025ztf,Cao:2025jwb,Kierkla:2025vwp,Huang:2025hos,Ning:2026nfs,Ai:2026zrs}.
Although these PBH bounds depend on the details of the formation mechanism and criterion, they conventionally exclude the region of relatively small $\beta/H$.

In this paper, we improve the formalism of Ref.~\cite{Azatov:2026sdm} and derive updated constraints on FOPT properties.
In particular, we quantify the impact of reheating on the temperature and baryon-asymmetry fluctuations.
In addition, we incorporate the scale dependence of the baryon-asymmetry power spectrum, using the transition-time power spectrum derived from finite-bubble statistics~\cite{Elor:2023xbz}.
We show that BBN disfavors the parameter region with
large $\alpha_{n}$ and small $\beta/H$.
However, no constraint is obtained for $\alpha_{n} < 0.1$.
These results highlight the importance of the reheating effect for BBN-based bounds on FOPTs.

Moreover, we show that the parameter region where gravitational waves (GWs) produced by FOPTs are compatible with the NANOGrav result~\cite{NANOGrav:2023gor, NANOGrav:2023hvm} may be disfavored by precise measurements of light element abundances if the power spectrum proposed in Ref.~\cite{Elor:2023xbz} remains valid down to sub-bubble scales.
In other words, the observed baryon asymmetry cannot simply predate the transition in such models.
The most robust way to evade the bound is baryogenesis after reheating.
Baryogenesis during the FOPT can also evade the bound.
This requires that the baryon number produced in each region scales with the local entropy injection.
In this case, the resulting $Y_{B}$ fluctuation is suppressed.
Our results thus suggest a way to constrain the timing of baryogenesis by combining GW signals and light-element abundances.

\section{Baryon asymmetry inhomogeneity by first-order phase transitions}
\label{sec:formalism}

The stochastic bubble nucleation rate plays a central role in describing FOPTs in the early universe.
In our analysis, we assume that the nucleation rate at the time $t$ is approximated by
\begin{align}
\label{eq:Gamma_t}
\Gamma(t) = \Gamma_{0} e^{ \beta (t - t_{n})} \,,
\end{align}
where $\beta$ and $t_{n}$ denote the inverse duration of the FOPT and the time at which bubble nucleation begins, respectively.
We define $t_{n}$ as the time at which the probability of nucleating one bubble per Hubble volume per Hubble time becomes of order unity, $\Gamma(t_{n})/H^4(t_{n}) = 1$, where $H(t)$ represents the Hubble parameter.
We denote the Hubble parameter and the temperature at $t_{n}$ by $H_{n} \equiv H(t_{n})$ and $T_{n}$, respectively.
As shown in Eq.~\eqref{eq:Gamma_t}, the bubble nucleation is exponentially enhanced after $t > t_{n}$.
To express the evolution of the FOPT, we introduce the fraction of the false vacuum region as
\begin{align}
F(t) = \exp \left[ - \int_{t_n}^{t} dt' \Gamma(t') V(t, t') \right] \equiv e^{-I(t)}\,,
\end{align}
with
\begin{align}
V(t, t') = \frac{4\pi}{3} a^3(t') \left[ \int_{t'}^{t} \frac{v_{w} d \tilde t}{a(\tilde t)}  \right]^3 \,,
\end{align}
where $a(t)$ and $v_{w}$ denote the scale factor and the wall velocity, respectively.
We set $v_{w} = 1$ throughout for simplicity.
The method used to compute the scale factor is described in Appendix~\ref{app:background}.

Since bubble nucleation is a stochastic process, it may be delayed over an extended region of space.
As a result, a large amount of vacuum energy can remain stored in the symmetric phase for an extended period.
Meanwhile, in regions where the transition has already completed, the released vacuum energy has been converted into radiation, whose density is further diluted by the cosmological expansion.
Consequently, a temperature inhomogeneity is expected to develop during a supercooled FOPT.
If the baryon asymmetry $\eta$ was already generated before the FOPT, this temperature inhomogeneity is directly inherited by $\eta$, since $\eta \propto n_{\gamma}^{-1} \sim T^{-3} $, where $n_{\gamma}$ denotes the number density of photons.
The inhomogeneity of the baryon asymmetry therefore encodes information about the FOPT.

The time evolution of the radiation energy density in a region where the FOPT completes at $t'$ is given by
\begin{align}
\rho_{R}(t,t') = \rho_{R}^{0}(t) + \rho_{V} \Theta(t - t')\left[ \frac{a(t')}{a(t)} \right]^4 \,,
\end{align}
where $\rho_{R}^{0}(t)$ denotes the radiation energy density that does not come from the FOPT, and $\rho_{V}$ represents the vacuum energy stored in the false vacuum domain.
Although the vacuum energy density could itself evolve in time if it depends on the scalar potential, we neglect this effect for simplicity.
In our analysis, we use the conventional definition of the latent heat $\alpha_n = \rho_{V}/\rho_{R}^{0}(t_{n})$ as an input parameter instead of $\rho_{V}$.
Throughout this paper, we assume that the released vacuum energy is efficiently converted into radiation of the Standard Model (SM) sector.
If a sizable fraction of it is instead released into a decoupled dark sector, the bound derived below is weakened, while the dark radiation is constrained by $\Delta N_{\rm eff}$~\cite{Planck:2018vyg}.

Regarding $\bar t$ as the average phase transition time, the radiation energy in the domain where the FOPT occurs at $\bar t$ obeys
\begin{align}
\label{eq:rho-ref}
\rho_{R}^{\rm ref}(t,\bar t) = \rho_{R}^{0}(t) + \rho_{V} \left[ \frac{a(\bar t)}{a(t)} \right]^4 \quad (t > \bar t)\,.
\end{align}
We regard $\rho_{R}^{\rm ref}(t)$ as the reference radiation energy.
On the other hand, in the region where the FOPT occurs at $t_{c} = \bar t + \delta t$, the time evolution of the radiation in the time range $t > t_{c}$ is described by
\begin{align}
\rho_{R}^{\rm dev}(t, t_{c}) = \rho_{R}^{0}(t) + \rho_{V} \left[ \frac{a(t_{c})}{a(t)} \right]^4 \,.
\end{align}
Then, the ratio of $\rho_{R}^{\rm ref}$ and $\rho_{R}^{\rm dev}$ at $t > \max(t_{c}, \bar t)$ can be expressed by
\begin{align}
\label{eq:rho-ratio}
\frac{\rho_{R}^{\rm dev}(t, t_{c})}{\rho_{R}^{\rm ref}(t, \bar t)}
= \frac{\rho_{R}^{0}(t) + \rho_{V} \left[ a(t_{c})/a(t) \right]^4}{\rho_{R}^{0}(t) + \rho_{V} \left[ a(\bar t)/a(t) \right]^4}
= \frac{g_{*}(T_{\rm dev})}{g_{*}(T_{\rm ref})} \left( \frac{T_{\rm dev}}{T_{\rm ref}} \right)^4 \,,
\end{align}
where $T_{\rm ref}$ and $T_{\rm dev}$ are the temperatures in the mean-transition region and early/late-transition regions, respectively.
The factor $g_{*}(T)$ represents the number of effective degrees of freedom for relativistic particles.
Equations~\eqref{eq:rho-ref}--\eqref{eq:rho-ratio} assume that each radiation component scales as $a^{-4}$ after the transition.
This scaling is exact only for a constant $g_{*}$.
We use these equations only to illustrate the origin of the temperature inhomogeneity.
In the numerical analysis, we instead follow the comoving entropy of each region with the local reheating map in Appendix~\ref{app:background}.
The two procedures coincide for a constant $g_{*}$.

To describe the variation of the reheating process, we introduce the probability distribution in terms of the local phase time $t_c$ as
\begin{align}
\label{eq:p_tc}
p(t_c) = - \frac{dF(t_{c})}{dt_{c}} ~~ \text{with} ~~ \int_{t_n}^{\infty} p(t_c) d t_c =  1 \,.
\end{align}
Using the probability distribution $p(t_c)$, we can define the time average of a quantity $X$ depending on $t_c$ as
\begin{align}
\braket{X} \equiv \int dt_c \,p(t_c) X(t_c) \,.
\end{align}
For the average phase transition time $\bar t$, it is defined by
\begin{align}
\bar t = \braket{t_{c}} = t_n + \int_{t_n}^{\infty} F(t) dt \,.
\end{align}

We assume that the baryon asymmetry is generated before the FOPT and that the comoving baryon number is conserved during the transition as
\begin{align}
N_B\equiv a^3(t)n_B(t)=\mathrm{const} \,.
\end{align}
To compare the mean-transition and early/late-transition regions on the same time hypersurface, we introduce
\begin{align}
t_f \equiv \max(t_c,\bar t) \,.
\end{align}
At $t=t_f$, both the reference region and the region undergoing its local transition at $t_c$ have completed their local conversion.
Then, at that time, their baryon number densities are identical before including baryon diffusion
\begin{align}
n_B^{\rm dev}(t_f) = n_B^{\rm ref}(t_f) \,.
\end{align}
Therefore, the baryon-to-entropy ratio $Y_B \equiv n_B/s_R$ differs between the two regions only through the local entropy density.
Using
\begin{align}
s_R(T)
=
\frac{2\pi^2}{45}
g_{*s}(T)T^3,
\end{align}
the ratio of $Y_{B}$ in the reference and early/late-transition regions can be given by
\begin{align}
\label{eq:YB-dev-ave}
\frac{
Y_B^{\rm dev}(t_f;t_c)
}{
Y_B^{\rm ref}(t_f;\bar t)
}
=
\frac{ g_{*s}(T_{\rm ref}) }{ g_{*s}(T_{\rm dev}) }
\left( \frac{T_{\rm ref}}{T_{\rm dev}} \right)^3 \,,
\end{align}
where $g_{*s}(T)$ is the effective number of degrees of freedom for the entropy at temperature $T$.
Since $Y_{B}$ in each region is conserved after its local transition, Eq.~\eqref{eq:YB-dev-ave} can also be evaluated right after the transition.
This gives $Y_B^{\rm dev}/Y_B^{\rm ref} = S_R(\bar t)/S_R(t_c)$.
Here, $S_R(t_c)$ is the comoving radiation entropy right after the transition at $t_c$, defined in Eq.~\eqref{eq:app-local-comoving-entropy}.
We use this expression in the numerical analysis.

Then, the fluctuation amplitude of the baryon inhomogeneity generated by the FOPT is defined by
\begin{align}
\label{eq:epsilon0-YB}
\epsilon_0^2
\left(
\alpha_n, \beta/H_n, T_n
\right)
\equiv
\frac{
\displaystyle
\Braket{\left(
Y_B^{\rm dev}/Y_B^{\rm ref}
\right)^2}
}{
\displaystyle
\braket{Y_B^{\rm dev}/Y_B^{\rm ref}}^2
}
-1 \,.
\end{align}
Since $Y_B^{\rm ref}$ is independent of $t_c$, this can be equivalently written as
\begin{align}
\label{eq:epsilon0}
\epsilon_0^2
\left(
\alpha_n, \beta/H_n, T_n
\right)
=
\frac{ \braket{ \left(Y_B^{\rm dev}\right)^2 } }{ \braket{Y_B^{\rm dev}}^2 } -1 \,.
\end{align}

In our framework, the baryon asymmetry inhomogeneity is defined by the baryon-to-entropy ratio $Y_{B}$, whereas the BBN bound derived in Ref.~\cite{Bagherian:2025puf} is given for the inhomogeneity of the baryon-to-photon ratio $\eta$.
Using the photon number density $n_{\gamma}(T) = 2\zeta(3)T^3/\pi^2$, the two quantities are related by
\begin{align}
\label{eq:eta-YB}
\eta = \frac{s_{R}(T)}{n_{\gamma}(T)}\, Y_{B} = \frac{\pi^4}{45\,\zeta(3)}\, g_{*s}(T)\, Y_{B} \,.
\end{align}
Thus, the ratio of $\eta$ in the early/late-transition region to that in the reference region is given by
\begin{align}
\label{eq:eta-ratio}
\frac{\eta^{\rm dev}}{\eta^{\rm ref}} = \frac{g_{*s}(T_{\rm dev})}{g_{*s}(T_{\rm ref})}\, \frac{Y_{B}^{\rm dev}}{Y_{B}^{\rm ref}} \,.
\end{align}
The fluctuation $\epsilon_{0}$ in Eq.~\eqref{eq:epsilon0} is evaluated right after the FOPT, i.e., at $t = t_{f}$, where the two regions have different temperatures.
At this time, the fluctuation of $\eta$ differs from that of $Y_{B}$ by the prefactor $g_{*s}(T_{\rm dev})/g_{*s}(T_{\rm ref})$ in Eq.~\eqref{eq:eta-ratio}, which deviates from unity when $g_{*s}$ varies between $T_{\rm dev}$ and $T_{\rm ref}$.
Moreover, the two quantities evolve differently after $t = t_{f}$.
$Y_{B}$ in each region is conserved apart from nucleon diffusion, because both the comoving baryon number and the comoving entropy are conserved.
In contrast, $\eta$ is not conserved when $g_{*s}$ changes, e.g., during the QCD crossover, because the comoving photon number changes accordingly.
Since regions with different temperatures experience the change of $g_{*s}$ at different times, the prefactor in Eq.~\eqref{eq:eta-ratio} changes in time, and it approaches unity once all regions reach a common temperature.
By the BBN epoch, the temperature inhomogeneity is expected to be smoothed out, while $Y_{B}$ of each fluid element remains unchanged because baryons and photons are tightly coupled~\cite{Bagherian:2025puf}.
Then, all regions share the same $g_{*s}$, and the fluctuation of $\eta$ relevant for BBN coincides with that of $Y_{B}$.
We note that this relation holds even if a small temperature fluctuation remains, since $g_{*s}$ is almost constant around $T \sim 1\,{\rm MeV}$ before the electron-positron annihilation.
Under the assumption of adiabatic evolution of each fluid element after local reheating, we use $Y_B$ as the fiducial matching variable. 
Once the temperature perturbation has relaxed and $s_R$ and $n_\gamma$ become spatially uniform, its relative fluctuation can be matched onto the baryon isocurvature relevant for BBN.
If one instead applies the BBN bound to the fluctuation of $\eta$ evaluated at $t = t_{f}$, the result differs from ours due to the prefactor in Eq.~\eqref{eq:eta-ratio}.
This difference is significant when $g_{*s}$ changes steeply between $T_{\rm dev}$ and $T_{\rm ref}$, in particular around the QCD crossover, whereas the two results coincide when $g_{*s}$ is constant in this temperature range, as shown in Appendix~\ref{app:baryon_to_photon}.

As discussed in Ref.~\cite{Bagherian:2025puf}, the baryon asymmetry fluctuation is damped by proton and neutron diffusion before the onset of BBN.
The scale dependence of this diffusion process is therefore essential for deriving constraints on FOPTs.
The one-point probability distribution $p(t_c)$ introduced above determines the expected initial variance $\epsilon_0^2$, but it does not by itself determine the spatial correlation of the local conversion time.
Additional information about the transition-time power spectrum is required.

To account for the scale dependence of this variance, we use the transition-time power spectrum discussed in Ref.~\cite{Elor:2023xbz}.
Introducing the time $t_{c}({\bm x})$ when the phase transition occurs at a point ${\bm x}$, the deviation in the phase transition time from the average value can be defined by
\begin{align}
\delta t_{c}({\bm x}) \equiv t_{c}({\bm x}) - \bar t \,.
\end{align}
Using the convention of Fourier transformation
\begin{align}
\delta t_{c}(\bm{k}) = \int d^3 \bm{x} e^{- i \bm{k} \cdot \bm{x}} \delta t_{c}(\bm{x}) \,,
\end{align}
the power spectrum of $\delta t_{c}({\bm x})$ is defined by~\cite{Cai:2024nln,Elor:2023xbz, Greene:2026gnw}
\begin{align}
\braket{\delta t_{c}(\bm{k})\delta t_{c}(\bm{k'})}
= (2\pi)^3 \delta^{(3)}(\bm{k} + \bm{k'}) P_{\delta t} (k) \,.
\end{align}
Then, the corresponding dimensionless power spectrum $\mathcal{P}_{\delta t}$ can be defined as~\cite{Elor:2023xbz}
\begin{align}
\label{eq:P_delta_t}
\mathcal{P}_{\delta t}(k) = \frac{k^3}{2\pi^2} \left[ \frac{\bar H}{\beta} \right]^2 \int d^3 {\bm r} e^{i {\bm k} \cdot {\bm r}} \beta^2 \braket{ \delta t_{c}({\bm x}) \delta t_{c}({\bm y})} \,,
\end{align}
where ${\bm r} = {\bm x} - {\bm y}$ and $\bar H \equiv H(\bar t)$.

A fully nonlinear calculation of the baryon asymmetry power spectrum would require unequal-point statistics of the nonlinear reheating mapping.
As an effective prescription, we assume that (i) Eq.~\eqref{eq:epsilon0} is an averaged result of the full nonlinear mapping and (ii) the spectral shape is determined by the transition-time spectrum in Eq.~\eqref{eq:P_delta_t}.
Then, the baryon-asymmetry power spectrum can be approximated by
\begin{align}
\label{eq:epsilon_k}
\Delta^{2}(k) = \epsilon_{0}^2(\alpha_n, \beta/H_{n}, T_{n}) \frac{\mathcal{P}_{\delta t}(k)}{\int d \ln k \mathcal{P}_{\delta t}(k)} \,.
\end{align}
We note that the coarse-graining result can be reproduced by performing the integral of $\Delta^{2}(k)$ in the whole $k$ region as
\begin{align}
\int d \ln k \, \Delta^{2}(k) = \epsilon_{0}^{2}(\alpha_{n}, \beta/H_{n}, T_{n}) \,.
\end{align}

We stress that Eq.~\eqref{eq:epsilon_k} is an effective parametrization rather than an exact prediction of the baryon power spectrum.
In Eq.~\eqref{eq:epsilon_k}, the nonlinear reheating dynamics is included in the total fluctuation amplitude.
However, the spectral shape is approximated by the transition-time fluctuation spectrum obtained from finite-bubble statistics~\cite{Elor:2023xbz}.
Although these contributions should ultimately be described by a unified framework, such an analysis is beyond the scope of the present work.

We also note that Eq.~\eqref{eq:epsilon_k} depends on the latent heat $\alpha_n$, in contrast to the result of Ref.~\cite{Azatov:2026sdm}.
This difference comes from the fact that the reheating effect is not included in the derivation of Ref.~\cite{Azatov:2026sdm}.

Defining the scale-dependent diffusion effect as $\mathcal{T}_{p}(k)$, the RMS at the BBN epoch can be expressed by
\begin{align}
\label{eq:epsilon_BBN}
\epsilon_{\rm BBN}^2 = \int_{0}^{k_{\rm cut}} d \ln k \, \Delta^{2}(k) \left| \mathcal{T}_{p}(a_{\rm B}(\bar t)k) \right|^2 \,.
\end{align}
In performing the numerical analysis, we use the result shown in Figure 2 in Ref.~\cite{Bagherian:2025puf} as the expression of $\mathcal{T}_{p}(k)$.
For the cutoff scale $k_{\rm cut}$, we mainly consider the following two cases:
\begin{align}
\label{eq:k_cut}
k_{\rm cut} = \begin{cases}
\infty \,, \\
\beta/\left[(8\pi)^{1/3} v_{w}\right] \equiv k_{\rm th} \,,
\end{cases}
\end{align}
where $k_{\rm th}$ corresponds to the inverse of the mean bubble separation at the FOPT.
As emphasized in Ref.~\cite{Elor:2023xbz}, in the sub-bubble range $k > k_{\rm th}$, effects of turbulence and magnetohydrodynamics may give a significant correction to the two-point function $\mathcal{P}_{\delta t}$.
Since the integrand of Eq.~\eqref{eq:epsilon_BBN} is non-negative, discarding these modes with the second cutoff choice gives a conservative constraint.
On the other hand, the first choice $k_{\rm cut} = \infty$ should be regarded as an optimistic case.
The factor $a_{\rm B}(\bar t)$ is included to match the difference in the normalization of the scale factor in Ref.~\cite{Elor:2023xbz} and that of Ref.~\cite{Bagherian:2025puf}.
Namely, $k$ in Eqs.~\eqref{eq:P_delta_t}--\eqref{eq:k_cut} denotes the physical wavenumber at $t = \bar t$, while the argument of $\mathcal{T}_{p}$ is the comoving wavenumber normalized by $a(T_{0} = 1\,{\rm MeV}) = 1$.
When the entropy is conserved in the visible sector after the FOPT, $a_{B}(\bar t)$ is given by
\begin{align}
\label{eq:aBt}
a_{B}(\bar t) = \frac{a(\bar t)}{a(t_{\rm RH})} \frac{T_{0}}{T_{\rm RH}} \left[ \frac{g_{*s}(T_{0})}{g_{*s}(T_{\rm RH})} \right]^{1/3} \,.
\end{align}
The method of numerical determination for the reheating temperature $T_{\rm RH}$ is explained in Appendix~\ref{app:background}.

According to the analysis in Ref.~\cite{Bagherian:2025puf}, the measurement of the deuterium abundance from quasar absorption lines gives the following bound at the $1 \sigma$ level
\begin{align}
\label{eq:epsilon_BBN_bound}
\epsilon_{\rm BBN} < 0.28 \,.
\end{align}
We take into account this bound in the subsequent discussion.

\section{Constraints from baryon asymmetry inhomogeneity}

In this section, we present our numerical results based on the formalism of the previous section.

\begin{figure}[t]
    \centering
    \includegraphics[width=0.65\linewidth]{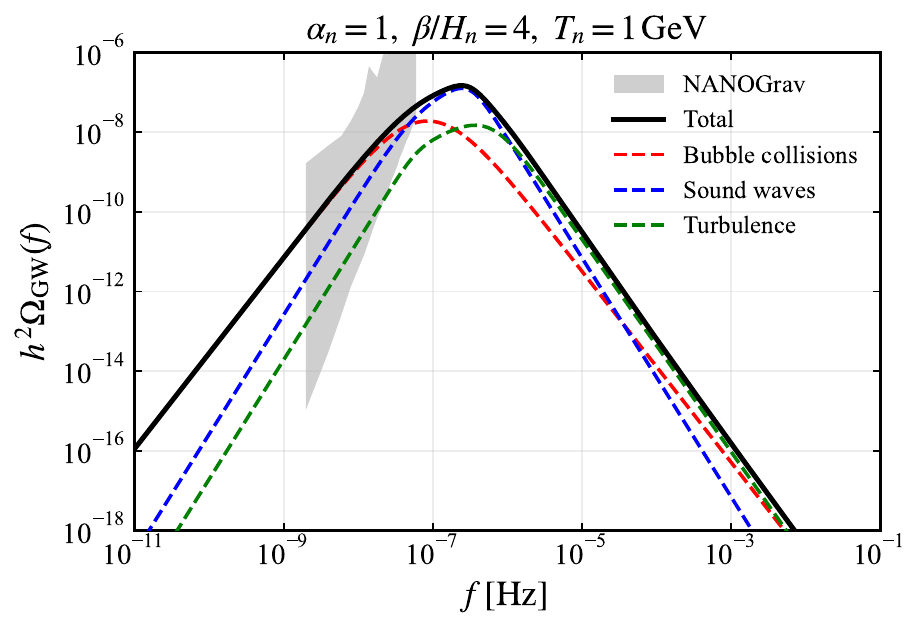}
    \caption{
    GW spectrum produced by the FOPT with $\alpha_{n} = 1$, $\beta/H_{n} =4$, and $T_{n} = 1\,{\rm GeV}$.
    The red, blue and green dashed lines are the GW contributions from bubble collisions, sound waves, and turbulence, respectively~\cite{Caprini:2024hue}.
    The black solid line denotes the total GW signal.
    The gray region represents the stochastic GW signal measured by NANOGrav~\cite{NANOGrav:2023gor, NANOGrav:2023hvm}.
    }
    \label{fig:GW_spectrum}
\end{figure}

\begin{figure*}[t]
    \centering
    \includegraphics[width=0.98\linewidth]{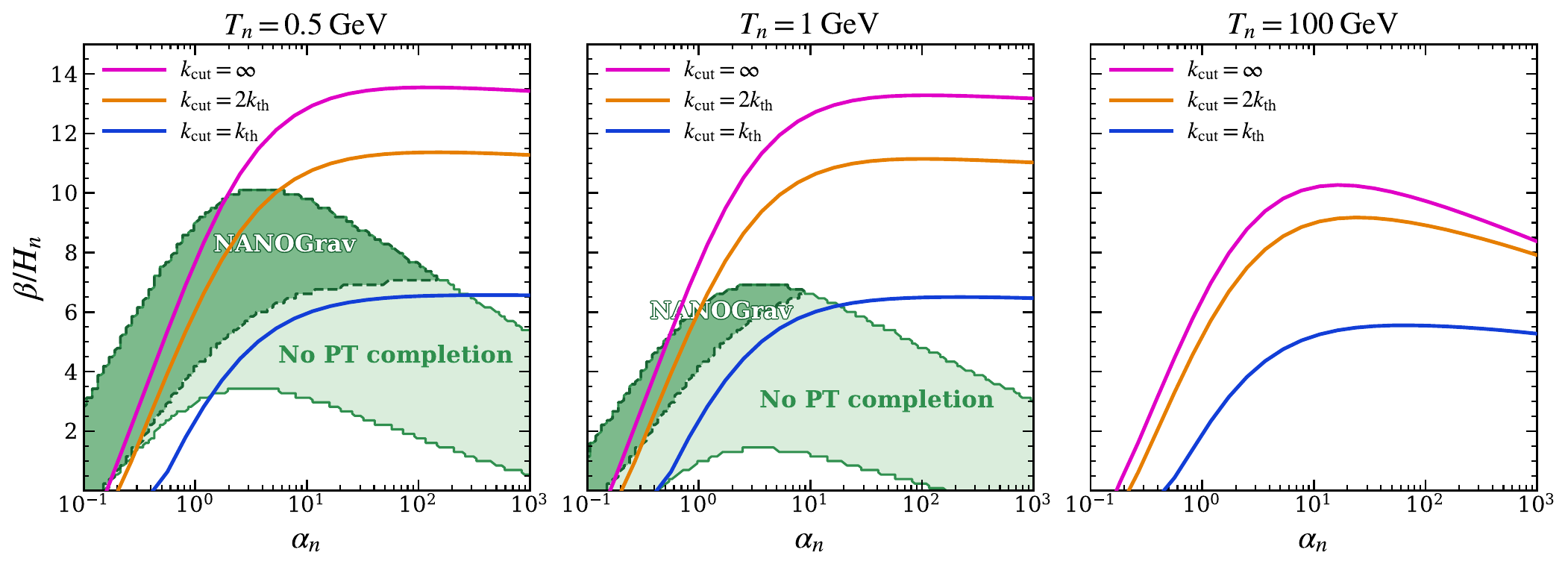}
    \caption{
    Constraints from the BBN measurement in the $(\alpha_{n}, \beta/H_n)$ plane for each nucleation temperature.
    The magenta and blue lines are the bounds obtained by combining Eq.~\eqref{eq:epsilon_BBN} with Eq.~\eqref{eq:epsilon_BBN_bound} for $k_{\rm cut} = \infty$ and $k_{\rm cut} = k_{\rm th}$ in Eq.~\eqref{eq:k_cut}, respectively.
    As an illustration, we also show the result with $k_{\rm cut} = 2 k_{\rm th}$ (orange line).
    The region below each line is disfavored at the $1\sigma$ level.
    In the dark green region, the GW spectrum is compatible with the NANOGrav result~\cite{NANOGrav:2023gor,NANOGrav:2023hvm}.
    For the light green region, the phase transition completion condition in Eq.~\eqref{eq:PT_completion} is not satisfied even though the GW spectrum can match the NANOGrav results.
    }
    \label{fig:BBNbound_T1_T100}
\end{figure*}

A FOPT in the early universe produces a characteristic gravitational-wave signal through the dynamics of the nucleated bubbles (see Ref.~\cite{Caprini:2024hue} and references therein).
Figure~\ref{fig:GW_spectrum} shows the predicted gravitational-wave (GW) spectrum for $\alpha_{n} = 1$, $\beta/H_{n} = 4$, and $T_{n} = 1\,{\rm GeV}$.
Here we use the fitting functions summarized in Ref.~\cite{Caprini:2024hue}.
The gray region denotes the stochastic GW signal measured by NANOGrav~\cite{NANOGrav:2023gor, NANOGrav:2023hvm}.
As we explain below, combining these gravitational-wave predictions with the BBN constraint can, in certain cases, constrain the timing of baryogenesis.

Figure~\ref{fig:BBNbound_T1_T100} shows constraints obtained by Eqs.~\eqref{eq:epsilon_BBN} and \eqref{eq:epsilon_BBN_bound} for each $k_{\rm cut}$ in Eq.~\eqref{eq:k_cut}.
We also show the parameter region in which the predicted GW spectrum matches the NANOGrav result while satisfying the requirement that the phase transition completes~\cite{Ellis:2018mja}.
We use the following condition as the criterion of phase transition completion
\begin{align}
\label{eq:PT_completion}
3 H(t_{p}) - \frac{d I(t_{p})}{dt} < 0 \,,
\end{align}
where $t_{p}$ denotes the percolation time defined by $F(t_{p}) = 0.71$.
The left and middle panels in Figure~\ref{fig:BBNbound_T1_T100} indicate that the BBN bound cannot exclude the darker green region when we use the conservative cutoff scale $k_{\rm cut} = k_{\rm th}$.
However, if we assume that the power spectrum in Eq.~\eqref{eq:epsilon_k} remains valid down to sub-bubble scales, the darker green region is largely disfavored by the BBN bound (the magenta line).
We note that a large part of the darker green region is still constrained even if we include only the modes with $k<2k_{\rm th}$.

Figure~\ref{fig:BBNbound_T1_T100} indicates that no bound from the baryon inhomogeneity is obtained unless $\alpha_{n}$ is sufficiently large: $\alpha_{n} \gtrsim 0.1$ for $k_{\rm cut} = \infty$ and $\alpha_{n} \gtrsim 0.5$ for $k_{\rm cut} = k_{\rm th}$.
This improves on the previous result of Ref.~\cite{Azatov:2026sdm}, where the bound on $\beta/H$ is independent of $\alpha_{n}$.
For large $\alpha_{n}$, we obtain the bound $\beta/H_{n} \lesssim 5\text{--}6$ for $k_{\rm cut} = k_{\rm th}$ and $\beta/H_{n} \lesssim 10\text{--}13$ for $k_{\rm cut} = \infty$ in the range $0.5\,{\rm GeV}<T_{n}<100\,{\rm GeV}$, consistent with the qualitative estimate of Ref.~\cite{Bagherian:2025puf}.

For the left and middle panels in Figure~\ref{fig:BBNbound_T1_T100}, the BBN bound with $k_{\rm cut} = \infty$ saturates at an almost constant value of $\beta/H_{n}$.
This is because the $\alpha_{n}$ dependence of $\epsilon_{0}^2(\alpha_{n}, \beta/H_{n}, T_{n})$ becomes modest once $\alpha_{n} \gg 1$.
In contrast, in the right panel of Figure~\ref{fig:BBNbound_T1_T100}, the BBN bound with $k_{\rm cut} = \infty$ decreases as $\alpha_{n}$ increases.
This difference comes from the effect of $a_{B}(\bar t)$ in Eq.~\eqref{eq:epsilon_BBN}.
When we consider the case with $\alpha_{n} \gg 1$, the reheating temperature $T_{\rm RH}$ is related to the nucleation temperature by $T_{\rm RH} \sim \alpha_{n}^{1/4} T_{n}$.
Substituting the relation into Eq.~\eqref{eq:aBt}, we obtain $a_{B}(\bar t) \propto \alpha_{n}^{-1/4}$.
Around the BBN epoch, the diffusion lengths of protons and neutrons are given respectively by $d_{p} \sim 10^{4}\,{\rm cm}$ and $d_{n} \sim 10^{6}\,{\rm cm}$~\cite{Bagherian:2025puf}.
On the other hand, in the radiation dominant universe with the temperature $T$, the comoving Hubble scale can be expressed by
\begin{align}
\frac{H^{-1}(T)}{a(T)}
\simeq 3 \times 10^{5}\,{\rm cm} \left( \frac{100\,{\rm GeV}}{T} \right) \left( \frac{106.75}{g_{*}(T)} \right)^{\frac{1}{6}} \,,
\end{align}
where we assumed $a(T_{0}=1\,{\rm MeV}) = 1$ and $g_{*}(T) = g_{*s}(T)$.
Therefore, comparing the horizon scale at $T_{n} =100\,{\rm GeV}$ with the proton diffusion length, we obtain $H^{-1}(T_n)/a(T_{n}) \sim 30\,d_{p}$.
This shows that proton diffusion remains ineffective for horizon-scale modes even by the time of BBN.
However, for fixed $\beta/H_{n}$, the inverse duration itself grows as $\beta \propto H_{n} \propto (1+\alpha_{n})^{1/2}$, because $H_{n}$ includes the vacuum energy.
Since $a_{B}(\bar t) k \sim a_{B}(\bar t) \beta \propto \alpha_{n}^{1/4}$ in the large $\alpha_{n}$ limit, proton diffusion becomes active and the resulting diffusion factor $\mathcal{T}_{p}(a_{B}(\bar t) k)$ leads to a large suppression.
As a result, the magenta line in the right panel of Figure~\ref{fig:BBNbound_T1_T100} decreases in the large $\alpha_{n}$ region.

\begin{figure}[t]
    \centering
    \includegraphics[width=0.65\linewidth]{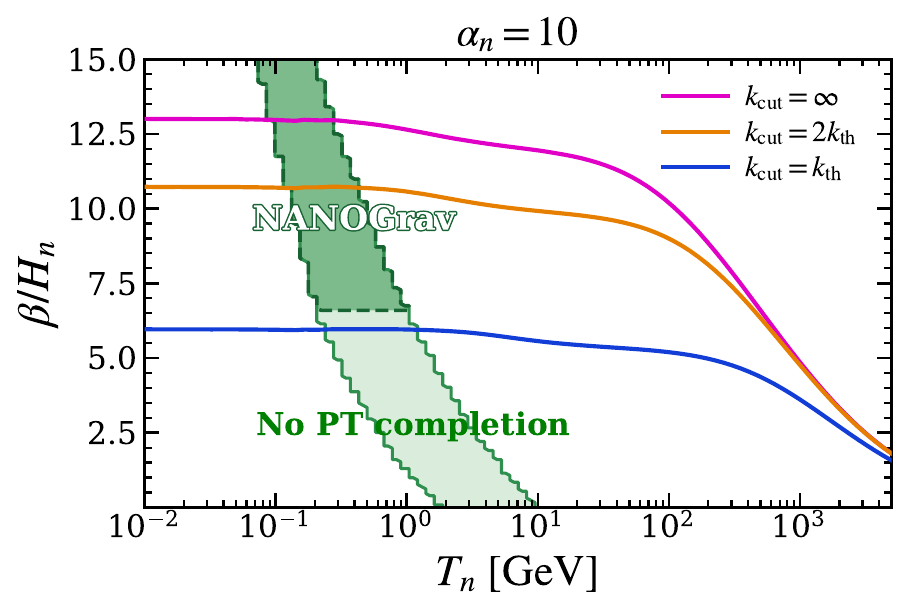}
    \caption{
    Constraint on the baryon-asymmetry inhomogeneity in the $(T_{n}, \beta/H_{n})$ plane for $\alpha_{n} = 10$.
    The definition of colored lines and regions is the same as Fig.~\ref{fig:BBNbound_T1_T100}.
    }
    \label{fig:BBN_fixed_alpha10}
\end{figure}

Figure~\ref{fig:BBN_fixed_alpha10} shows the parameter region constrained by the baryon-asymmetry inhomogeneity in the $(T_{n}, \beta/H_{n})$ plane for $\alpha_{n} = 10$.
As the value of $T_{n}$ increases, the BBN bound weakens, because the inhomogeneity scale set by the Hubble scale shrinks relative to the diffusion length.
The behavior of the magenta and blue lines is consistent with the qualitative expectation discussed in Ref.~\cite{Bagherian:2025puf}.

We emphasize that Figs.~\ref{fig:BBNbound_T1_T100} and \ref{fig:BBN_fixed_alpha10} assume that baryogenesis is already completed before the FOPT.
If baryogenesis instead occurs homogeneously after reheating, models in the darker green region below the magenta line can evade the BBN bound (e.g., Ref.~\cite{Fujikura:2024jto}).
On the other hand, if the baryon asymmetry is produced during the FOPT, e.g., at the bubble walls, the spatial variation of the local baryon production induces its own correlated fluctuation of $Y_{B}$, and the bound is not evaded in general.
In other words, interpreting the NANOGrav signal as evidence for a FOPT in such models implies that the observed baryon asymmetry cannot predate the transition, and should be produced after reheating.
Therefore, by combining GW observations with the BBN constraint based on the baryon asymmetry inhomogeneity, we may be able to constrain the timing of baryogenesis.

\section{Discussion and Conclusion}

We have examined how precision measurements of the light elements produced during BBN can constrain the properties of FOPTs in the early universe.
Assuming (i) the pre-existing baryon asymmetry before the FOPT, (ii) new physics sectors inducing the FOPT sufficiently couple with the SM sector, and (iii) the inhomogeneity of $Y_{B}$ is characterized by the transition-time fluctuation spectrum based on finite-bubble statistics~\cite{Elor:2023xbz}, we obtained that BBN measurements disfavor at the $1\sigma$ level the parameter region with $\beta/H_{n} \lesssim 5\text{--}6$ for $\alpha_{n} \gtrsim 0.5$ with the conservative cutoff $k_{\rm cut} = k_{\rm th}$, and $\beta/H_{n} \lesssim 10\text{--}13$ for $\alpha_{n} \gtrsim 0.1$ with $k_{\rm cut} = \infty$, in the temperature range $0.5\,{\rm GeV}< T_{n} < 100\,{\rm GeV}$.
In addition, we have shown that the parameter region where the predicted GW spectra are compatible with the NANOGrav result could be disfavored by the BBN-based bound if the power spectrum in Eq.~\eqref{eq:epsilon_k} remains valid down to sub-bubble scales.
To avoid the BBN bound in new physics models within that parameter region, the baryon asymmetry cannot simply be produced before the FOPT.
Instead, baryogenesis should proceed after reheating.
Alternatively, baryogenesis may proceed during the FOPT if the local baryon production tracks the local entropy injection.
Therefore, under the assumptions summarized above, our results imply that the timing of baryogenesis can be constrained by measuring GW signals and light-element abundances.

Finally, we comment on several potential uncertainties that could affect our results.
First, in deriving the constraint from precision deuterium measurements, we used Eq.~\eqref{eq:epsilon_BBN} to account for the contribution of baryon diffusion.
Then, we have considered the result with each cutoff scale shown in Eq.~\eqref{eq:k_cut}.
As shown in Fig.~\ref{fig:BBNbound_T1_T100} and Fig.~\ref{fig:BBN_fixed_alpha10}, the dependence of the cutoff scale can drastically change the constrained region.
This indicates that a complete derivation of the power spectrum for the baryon-to-entropy ratio $Y_{B}$ is necessary to precisely determine the BBN-based bound.
We leave this significant analysis for future work.

Second, in our analysis, the bubble nucleation is approximated by the simple exponential expression in Eq.~\eqref{eq:Gamma_t}.
However, since the BBN-based bound we obtained disfavors the parameter region with small $\beta/H_{n}$, the quadratic term in the power of exponent in the bubble nucleation rate can play an important role, i.e., $\Gamma = \Gamma_{0} e^{\beta (t - t_{n}) - \zeta^2(t - t_{n})^2/2}$~\cite{Megevand:2016lpr, Lewicki:2023ioy, Kanemura:2024pae}.
However, when we use the improved tunneling rate, the transition-time fluctuation spectrum discussed in Ref.~\cite{Elor:2023xbz} should be modified in general.
This is beyond the scope of the present paper.

\paragraph{Note added}

During the final preparation of this paper, Ref.~\cite{Elahi:2026} appeared.
It also studied BBN constraints on the baryon isocurvature induced by inhomogeneous reheating from FOPTs and domain walls.
Using a linearized estimate, they obtained the bound $\beta/H_{\star} \gtrsim 10.3\,\alpha/(1+\alpha)$, which is qualitatively consistent with our results.
Our analysis complements theirs by using the nonlinear local reheating map and by including the scale dependence of the transition-time spectrum and proton diffusion.
Ref.~\cite{Elahi:2026} assumes that the isocurvature survives until BBN for reheating temperatures below $\sim 3\,{\rm TeV}$, based on the criterion for horizon-scale modes in Ref.~\cite{Bagherian:2025puf}.
However, the modes sourced by the FOPT are sub-horizon, and proton diffusion weakens the bound already for $T_{n} \gtrsim 10^{2}\,{\rm GeV}$, as shown in Fig.~\ref{fig:BBN_fixed_alpha10}.
Therefore, the BBN bound on FOPTs depends on the transition temperature well below $3\,{\rm TeV}$, although this does not affect the PTA-relevant temperatures.

\section*{Acknowledgement}

I would like to thank Jia Liu for helpful and fruitful discussions.

\appendix

\section{Numerical background evolution}
\label{app:background}

In this appendix, we describe the method used to compute the scale factor and other quantities relevant to reheating.

To simplify the subsequent formula, we define the following dimensionless quantities
\begin{align}
x \equiv H_{n} (t - t_{n}) \,, ~
R(x) \equiv \frac{\bar \rho_R(t)}{\rho_R^0(t_n)} \,, ~
h(x) \equiv \frac{H(t)}{H_n} \,.
\label{eq:app-background-variables}
\end{align}
These quantities satisfy the following normalization
\begin{align}
R(0)=F(0)=1 \,.
\end{align}
Here, $\bar \rho_R$ denotes the volume-averaged radiation energy density.
While the normalization $a(t_n)=1$ is used only in solving the background equations, the BBN normalization is restored through Eq.~\eqref{eq:aBt}.

The time evolution of the scale factor and the radiation energy density are determined from the coupled equations
\begin{align}
& \frac{1}{a}\frac{da}{dx} = h \,,
\label{eq:app-scale-factor}
\\
&\frac{dR}{dx}
+
3h\left(R+P\right) = -\alpha_n\frac{dF(x)}{dx} \,,
\label{eq:app-radiation-background}
\\
& h^2(x) = \frac{R+\alpha_n F(x)}{1+\alpha_n} \,,
\label{eq:app-friedmann}
\end{align}
where
\begin{align}
P(x)
\equiv
\frac{p_R(T)}{\rho_R^0(t_n)}.
\end{align}
The temperature corresponding to $R(x)$ is obtained from
\begin{align}
R(x)
=
\frac{g_*(T)}{g_*(T_n)}
\left(\frac{T}{T_n}\right)^4,
\label{eq:app-temperature-from-radiation}
\end{align}
and the radiation pressure is evaluated using
\begin{align}
p_R(T)
=
T s_R(T)-\rho_R(T) \,.
\label{eq:app-radiation-pressure}
\end{align}
In the numerical analysis, we use the energy-density column of Ref.~\cite{Husdal:2016haj} for both $g_{*}$ and $g_{*s}$.
With this choice, Eq.~\eqref{eq:app-radiation-pressure} reduces to $p_R = \rho_R/3$.

The source term on the right-hand side of Eq.~\eqref{eq:app-radiation-background} describes the conversion of vacuum energy into the visible radiation energy density.
Since $dF/dx<0$, this term gives a positive contribution to $\bar \rho_{R}$.
The same background solution $a(x)$, $R(x)$, and $F(x)$ is used to determine the transition-time distribution, the average transition time, the local reheating map, and the reheating temperature.
We do not introduce a separate homogeneous thermalization background.

To evaluate the local reheating effect, we first determine the radiation temperature $T_{\rm pre}(t_c)$ just before the phase transition occurs.
The unconverted radiation evolves adiabatically along the common background scale factor and therefore satisfies the following relation
\begin{align}
a(t_c) \, T_{\rm pre}(t_c)\, g_{*s}^{1/3} \left(T_{\rm pre}(t_c)\right)
= a(t_n)\, T_n\, g_{*s}^{1/3}(T_n) \,.
\label{eq:app-pre-transition-temperature}
\end{align}
Assuming the rapid conversion of the local vacuum energy into radiation, the temperature $T_{\rm post}(t_c)$ immediately after the phase transition is determined by
\begin{align}
g_{*}\left(T_{\rm post}\right)T_{\rm post}^4
= g_{*}\left(T_{\rm pre}\right)T_{\rm pre}^4 + \alpha_n g_{*}(T_n) T_n^4 \,.
\label{eq:app-local-reheating}
\end{align}

The comoving radiation entropy produced after the phase transition at $t_c$ is proportional to
\begin{align}
S_R(t_c) \equiv a^3(t_c)s_R\!\left(T_{\rm post}(t_c)\right) \,.
\label{eq:app-local-comoving-entropy}
\end{align}
Since we assume that the comoving baryon number is conserved, the baryon-to-entropy ratio in that domain satisfies $Y_B(t_c) \propto 1/S_R(t_c)$.
The nonlinear one-point variance $\epsilon_0^2$ in the main text is evaluated by averaging $Y_B(t_c)$ over the probability distribution in Eq.~\eqref{eq:p_tc}.

Finally, we describe the reheating time $t_{\rm RH}$ and the corresponding temperature $T_{\rm RH}$ used in our analysis.
Defining the reheating time as $F(t_{\rm RH})\simeq 10^{-4}$ and introducing $R_{\rm RH} \equiv R \left( H_n(t_{\rm RH}-t_n)\right)$, the reheating temperature $T_{\rm RH}$ is determined by
\begin{align}
g_{*}(T_{\rm RH})T_{\rm RH}^4 = R_{\rm RH} \, g_{*}(T_{n})T_{n}^4 \,.
\label{eq:app-reheating-temperature}
\end{align}
The energy released from the false vacuum is already included in $R_{\rm RH}$ through the source term in Eq.~\eqref{eq:app-radiation-background}; it is therefore not added separately in Eq.~\eqref{eq:app-reheating-temperature}.

After $t_{\rm RH}$, we assume entropy conservation in the visible sector, so that the ratio $a(\bar t)/a(t_{\rm RH})$ appearing in Eq.~\eqref{eq:aBt} follows directly from the common solution of Eq.~\eqref{eq:app-scale-factor}.
Combining this ratio with the entropy-conserving evolution after reheating gives the scale factor $a_B(\bar t)$ in Eq.~\eqref{eq:aBt}.

\section{Framework with the baryon-to-photon ratio}
\label{app:baryon_to_photon}

In this appendix, we consider the difference in the constrained region using BBN observations.

\begin{figure}[t]
    \centering
    \includegraphics[width=0.65\linewidth]{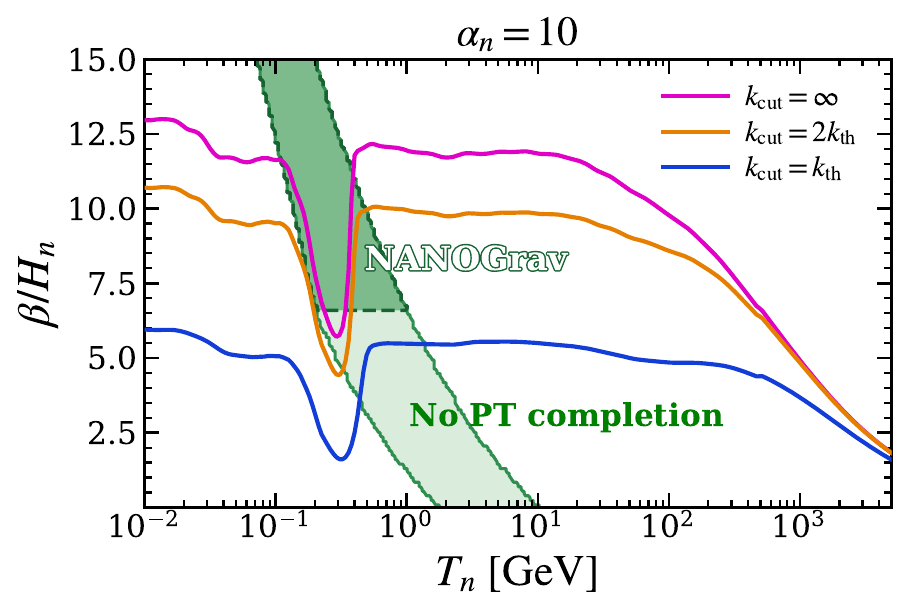}
    \caption{
    Constraint on the baryon-asymmetry inhomogeneity in the $(T_{n}, \beta/H_{n})$ plane for $\alpha_{n} = 10$ with the $\eta$ framework.
    The definition of colored lines and regions is the same as Fig.~\ref{fig:BBNbound_T1_T100}.
    \label{fig:alpha_beta_eta_scheme}
    }
\end{figure}

In the main text, we used the baryon-to-entropy ratio $Y_{B}$ to evaluate the fluctuation right after the FOPT.
Here, we instead define the fluctuation with the baryon-to-photon ratio $\eta$ evaluated at $t = t_{f}$, and compare the resulting bound with ours.

Figure~\ref{fig:alpha_beta_eta_scheme} shows the nucleation temperature dependence of the BBN bound with the $\eta$ framework.
We here use the following definition for $\epsilon_{0}$:
\begin{align}
\label{eq:epsilon0-eta}
\epsilon_0^2
\left(
\alpha_n, \beta/H_n, T_n
\right)
\equiv
\frac{
\displaystyle
\Braket{\left(
\eta^{\rm dev}/\eta^{\rm ref}
\right)^2}
}{
\displaystyle
\braket{\eta^{\rm dev}/\eta^{\rm ref}}^2
}
-1 \,,
\end{align}
where $\eta^{\rm dev}/\eta^{\rm ref}$ is evaluated at $t = t_{f}$ and is related to $Y_{B}^{\rm dev}/Y_{B}^{\rm ref}$ by Eq.~\eqref{eq:eta-ratio}.
Comparing Fig.~\ref{fig:BBN_fixed_alpha10} with Fig.~\ref{fig:alpha_beta_eta_scheme}, the two results coincide at low $T_{n}$, where $g_{*s}$ is almost constant, while the bound changes drastically around $T_{n} \sim 200\,{\rm MeV}$.
This change comes from the steep increase of $g_{*s}$ during the QCD crossover~\cite{Husdal:2016haj}, which makes the prefactor in Eq.~\eqref{eq:eta-ratio} deviate significantly from unity.
As explained in Section~\ref{sec:formalism}, this prefactor disappears by the BBN epoch, and hence Fig.~\ref{fig:BBN_fixed_alpha10} gives the physical bound.

\bibliographystyle{JHEP}
\bibliography{reference}

\end{document}